\documentclass[prl,10pt, a4paper, twocolumn]{revtex4}

\usepackage{amsmath}            
\usepackage{amsfonts}	
\usepackage[dvips]{graphicx, color}

\newcommand{\dero}[2]{\frac{\partial #1}{\partial #2}}

\begin{document}

\title{Random quantum Ising chains with competing interactions}

\author{David Carpentier}
\affiliation{Laboratoire de Physique de l'Ecole Normale
Sup{\'e}rieure de Lyon, 
46, All{\'e}e d'Italie, 69007 Lyon, France}
\author{Kay-Uwe Giering}
\affiliation{Laboratoire de Physique de l'Ecole Normale
Sup{\'e}rieure de Lyon, 
46, All{\'e}e d'Italie, 69007 Lyon, France}
\affiliation{
Fakult{\"a}t f{\"u}r Physik und Geowissenschaften, Universit{\"a}t Leipzig,
Augustusplatz 10, 04109 Leipzig, Germany}
\author{Pierre Pujol}
\affiliation{Laboratoire de Physique de l'Ecole Normale
Sup{\'e}rieure de Lyon, 
46, All{\'e}e d'Italie, 69007 Lyon, France}

\date{22 Juin 2005}

\begin{abstract}
In this paper we discuss the criticality of a quantum Ising spin chain with competing
random ferromagnetic and antiferromagnetic couplings. Quantum fluctuations are introduced {\it via} 
random local transverse fields. 
First we consider the chain with couplings between first and second 
neighbors only and then generalize the study to a quantum analog of the  
Viana-Bray model, defined on a small world random lattice. 
 We use the Dasgupta-Ma decimation technique, both analytically and numerically, and
 focus  on the scaling of the lattice topology, whose determination is necessary to define 
 any infinite disorder transition beyond the chain.  
In the first case, at the  transition the model renormalizes towards the chain, 
 with the infinite disorder fixed point described by Fisher. This 
corresponds to the irrelevance of the competition induced by the second neighbors couplings. 
 As opposed to this case, this infinite disorder transition is found to be unstable towards the introduction 
 of an arbitrary small density of long range couplings in the small world models. 
\end{abstract}

\maketitle

 Quantum fluctuations play a crucial role in 
 the spin glass phases of the Sr-doped cuprate La$_{2}$CuO$_{4}$ \cite{Kastner1998}, or the dipolar glass 
LiHo$_{x}$Y$_{1-x}$F$_{4}$ in a transverse field \cite{Aeppli1987}. Randomly coupled quantum 
two level systems also appear in the understanding of the dielectric response of low temperature 
amorphous solids \cite{Ludwig2003}, and as the main low frequency source of decoherence of 
solid state quantum bits \cite{Galperin2003}. In all cases, the quantum fluctuations compete with the 
random couplings between the spin, and tend to disorder the corresponding random ordered phases.  

 One of the simplest random quantum model to study this competition is probably the random Ising spin model in 
 a transverse magnetic field. 
\begin{equation}
\label{eq:randomIsing} 
H=-\sum_{i,j} J_{ij}  \sigma_i^z \sigma_{i+1}^z 
-\sum_{i} h_i \sigma_i^x , 
\end{equation} 
 where $\sigma_{i}^{x}, \sigma_{i}^{z}$ are the usual Pauli matrices, and the transverse fields $h_{i}$ 
 are responsible for the quantum tunneling fluctuations between the up and down states of the Ising spins.   
In a pioneering work, Fisher has given asymptotically  exact results 
for this random quantum Ising model with first neighbors 
random ferromagnetic bonds in one dimension \cite{Fisher1995}.
By using a decimation technique developed by Dasgupta and Ma \cite{Dasgupta1980},
he described the {\it infinite disorder} quantum phase transition of this model. 
Some of the main features of this peculiar transition were a diverging  
dynamical exponent $z$, and very strong inhomogeneities manifesting  through drastically different behavior 
between average and typical correlation functions. 

Natural extensions of these results to higher dimensions have proved to be difficult. In particular, 
an analytical implementation of the Ma-Dasgupta decimation beyond the simple chain is extremely
cumbersome. The reason is that any initial lattice except the chain is quickly randomized by the decimation. 
Thus one has to resort to a numerical implementation of this decimation \cite{Motrunich2000,Rieger2000}. 
 For two-dimensional regular lattices, the results for the random ferromagnetic Ising model 
 are consistent with the survival of an infinite disorder quantum 
 phase transition, albeit with exponents different from the one-dimensional case \cite{Motrunich2000}. 
 On the other hand, the quantum Ising spin glass, corresponding to the model (\ref{eq:randomIsing}) with 
 both ferromagnetic (positive) and antiferromagnetic (negative) couplings, was also studied in two and three dimensions 
 {\it via} Monte-Carlo simulations \cite{Rieger1996}. The numerical works found 
 no sign of an infinite disorder quantum critical point. Since the results for the random ferromagnetic model are
  expected to extends to the quantum Ising spin glass, this discrepancy 
  certainly deserves further work.  

In this perspective, we investigate in this paper  the stability of infinite disorder fixed point of the quantum 
Ising spin glass chain with respect to 
competing further neighbors couplings in two extreme cases. 
 In a first step, we focus in the case where second neighbors couplings are present in model (\ref{eq:randomIsing}) 
 besides the first neighbors couplings. The couplings are taken as either ferromagnetic or antiferromagnetic. 
By combining analytical (for small second neighbors couplings) and numerical decimation techniques 
we investigate the relevance of the presence of higher range couplings and their interplay with random signs in the couplings. 
 We pay a special attention to the topology of the renormalized lattice, which appears crucial in the precise 
 characterization of infinite disorder transitions. 

 A natural complement to this first case consists in 
 considering this quantum Ising spin glass on a random network, obtained by adding a finite 
 density of long range couplings between the chain's sites. Indeed, our model can be considered as a quantum analog of  
the classical spin models of Viana and Bray\cite{Viana1985}, although we keep a local regular topology besides
 the random long range couplings in our {\it small world} lattice\cite{Albert2002}.   
 In the case of classical spin glasses, these 
 random lattice models are a natural extrapolation between the short-range model and its mean-field  
 version. They
 undergo a finite temperature transition of  the mean-field type, albeit with peculiarities induced by the finite 
 connectivity\cite{Nikoletopoulos}. Wether the infinite disorder physics survives to this tendency towards mean-field like 
 physics is the natural question we will consider. 
 
In the first part of this letter, we consider the model (\ref{eq:randomIsing}) on a chain, with only 
first and second neighbors couplings (Zig-Zag ladder) \cite{footnote2}. 
Both the first and second neighbors couplings
$J_{i, i+1}^{(1)},J_{i, i+2}^{(2)}$ can be antiferromagnetic ($<0$) with probability 
$p$, and ferromagnetic with probability $1-p$. 
 The $|J_{i, i+1}^{(1)}|$ are uniformly distributed 
 between $0$ and $1$,  the $|J_{i, i+2}^{(2)}|$ between $0$ and $J^{(2)}_{max}$, and the transverse fields  $h_{i}$ 
 between $0$ and $h_{max}$.  
  Note that {\it via} an appropriate unitary transformation we can map this system onto one 
  where only the second neighbors couplings can have both signs, 
  but at the cost of a modification of the magnetic properties of the system. 
  Hence for clarity, we prefer to consider only the more natural choice 
 defined above.  
   
We will analyze the low temperatures behavior of this system by means of the
Dasgupta-Ma decimation technique \cite{Dasgupta1980} 
which was exploited by Fisher \cite{Fisher1995} in the case,
among others, of the  random ferromagnetic quantum Ising chain. 
Its extension to the present case of mixed coupling (anti-ferromagnetic and ferromagnetic) contains one 
supplementary rule as detailed below. The running energy scale $\Omega$ is defined  as the maximum of the amplitudes of bonds 
$|J_{ij}|$ and fields $h_i$. At each decimation step, if this maximum corresponds to a field $h_i$, the corresponding spin is 
frozen in the $x$ direction, generating new couplings  $\tilde{J}_{jk} = J_{jk} + (J_{ij}J_{ik} / \Omega)$ between 
all pairs $(j,k)$ previously connected with the spin $i$.    On the other hand, if the maximum is a ferromagnetic coupling $J_{ij}$, 
the two spins $i$ and $j$ are paired to form a new cluster $[ij]$ of magnetization $\mu_{[ij]}=\mu_i+\mu_j$ (where $\mu_i$ corresponds 
to the magnetization of cluster $i$), and coupling with site $k$ $J_{[ij]k}= J_{ik} + J_{jk}$ \cite{Fisher1995}. 
The new rule occurs when this maximum 
coupling is anti-ferromagnetic. In this case, if {\it e.g} the magnetization $\mu_i$ is larger than $\mu_j$, then the new cluster's 
magnetization reads $\mu_{[ij]}=\mu_i-\mu_j$, and the interaction with a third spin $k$ is $J_{[ij]k}= J_{ik} - J_{jk}$. In both 
cases, the effective transverse field acting on the new cluster is $h_{[ij]}=h_ih_j/\Omega$. 


 An analytical study of the scaling behavior of the model (\ref{eq:randomIsing}) under the above decimation rules  is 
 difficult even on the Zig-Zag ladder we consider. 
 As mentioned in the introduction,  couplings $J_{ij}$ are quickly generated on many length scales $|i-j|$, and the initial lattice is 
 quickly randomized (see below). 
  To fix the notation and clarify the procedure, it is useful to start by considering the evolution under the RG of the first neighbors chain, 
  extending the result of Ref. \onlinecite{Fisher1995} to the presence of anti-ferromagnetic couplings. 
We introduce the convenient logarithmic variables 
$\beta_{i} =\ln (\Omega/h_{i})$, $\zeta_{i,i+1}=\ln (\Omega/|J_{i,i+1}|)$ and  scaling parameter 
$\Gamma := \ln ( \Omega_0/\Omega )$ where $\Omega_{0}$ is the initial value of $\Omega$. Their 
``distributions'' densities are defined as $\mathcal{R}(\beta,\Gamma)$ for the fields, 
$\mathcal{P}^{(1+)}(\zeta,\Gamma)$ for the ferromagnetic bonds, 
and $\mathcal{P}^{(1-)}(\zeta,\Gamma)$ for the anti-ferromagnetic bonds. Note that while 
$\mathcal{R}(\beta,\Gamma)$ is normalized,  for the bonds only the sum 
$\mathcal{P}^{(1)}(\zeta,\Gamma)=\mathcal{P}^{(1+)}(\zeta,\Gamma)+\mathcal{P}^{(1-)}(\zeta,\Gamma)$ has 
a norm one. As can be deduced by a gauge transformation of (\ref{eq:randomIsing}), 
$\mathcal{R}(\beta,\Gamma)$ and $\mathcal{P}^{(1)}(\zeta,\Gamma)$ 
satisfy the same differential scaling equations than in the ferromagnetic case \cite{Fisher1995} provided we 
use the maximum instead of the sum in the above decimation rules, which is valid for broad enough distributions. 
  Finally, the function 
 $\mathcal{D}(\beta,\Gamma)=\mathcal{P}^{(1+)}(\zeta,\Gamma)-\mathcal{P}^{(1-)}(\zeta,\Gamma)$ 
 is found to satisfy the same scaling equation than $\mathcal{P}^{(1)}$. 
   The fixed point  
$\mathcal{R} =  \mathcal{P}^{(1)}=\mathcal{P}^*(x,\Gamma)=e^{-x/\Gamma}/\Gamma$ 
of the ferromagnetic chain\cite{Fisher1995}
is easily extended to the two following case : the above ferromagnetic point now corresponds to 
the solution $\mathcal{D}=\mathcal{P}^{*}$, or 
$\mathcal{P}^{(1+)} =\mathcal{P}^{(1)}= \mathcal{P}^{*}$, $\mathcal{P}^{(1-)}=0$. 
As expected,  it can be explicitely checked in the RG equations that 
this fixed point is unstable towards the proliferation of anti-ferromagnetic bonds. 
The new transition point corresponds to the solution $\mathcal{D}=0$, or 
$\mathcal{P}^{(1+)} =\mathcal{P}^{(1-)}= \mathcal{P}^{*}/2$, 
corresponding to an equal density of random positive and negative couplings. Hence, we will loosely call it 
 the spin glass fixed point by analogy with the physics of the classical model in higher dimensions. 
The  characteristics of the transtion from the  the ferromagnetic to the disordered phase obtained by Fisher \cite{Fisher1995} translate 
to the present spin-glass fixed point into an average linear susceptibility (under the application 
 of a small $z$ field $\tilde{h}$) which diverges 
 as $\chi(T) \sim |\ln T|^{\phi - 2} / T$ where  $ \phi = (1+\sqrt{5})/2$ . Similarly, we extract 
 the scaling behaviour of the average non linear susceptibility 
 $\chi_{nl}(T)=\left[ \left.\dero{{}^3}{\tilde{h}^3}\right|_{\tilde{h}=0} <M> (\tilde{h}) \right] $, where 
 $[\dots ]$ denotes an ensemble average, as 
$\chi_{nl}(T) \sim |\ln T|^{2\phi - 2} / T^3$. 
%


 Having clearly defined the notation and fixed points for the chain, we can now study perturbatively their stability  with respect to small second neighbors competing interactions. 
To first order, such an analysis can be conducted by considering  the presence of  $J^{(2)}$ negligible 
compared to the $J^{(1)}$, and checking whether this condition is self-consistently preserved under the re-scaling. 
 More precisely, we will assume that 
{\it (i)} a $J_{i, i+2}^{(2)} $ will never constitute the
highest energy in the system and therefore never be decimated
{\it (ii)} in sums, the $J_{i, i+2}^{(2)}$ are negligible with
respect to $J_{i, i+1}^{(1)}$ 
{\it (iii)} creation of third neighbour couplings out of second 
neighbour couplings can be neglected. 
 As above, we define the distribution 
 $\mathcal{P}^{(2)}(\zeta,\Gamma):=\mathcal{P}^{(2+)}(\zeta, \Gamma) + \mathcal{P}^{(2-)}(\zeta,
\Gamma)$ as the sum of ``distributions'' of positive and negative  next nearest neighbour
couplings. With the above hypothesis, its scaling behavior is found to be described by 
\begin{align} 
&\dero{\mathcal{P}^{(2)}(\zeta)}{\Gamma}  = 
\dero{\mathcal{P}^{(2)}(\zeta)}{\zeta} 
- \mathcal{P}^{(2)}(\zeta)
\left( 2 \mathcal{R}(0) + \mathcal{P}^{(1)}(0)  \right)  \nonumber\\
&+ 2 \mathcal{R}(0) \int_0^{\infty} \mathrm{d} \zeta_1\
\mathrm{d} \zeta_2\ \mathcal{P}^{(1)}(\zeta_1) \mathcal{P}^{(2)}(\zeta_2)  
 \delta (\zeta - \zeta_1 - \zeta_2) \nonumber\\
 &  + \mathcal{P}^{(1)}(0)  \
\delta (\zeta - \Lambda)
\label{eq:RG-J2} 
\end{align}
where $\Lambda$ is an arbitrary large constant which stands for the negligible $J$ in log. coordinates, and  
is taken to $\infty$ at the end of calculations. The $\Gamma$ 
dependance of the distribution has been omitted for clarity. With the above hypothesis, the
probability distributions for fields and nearest neighbour couplings still follow the equations for the chain. 
Hence, at the ``Spin Glass'' critical point, 
we can  insert the scaling form $\mathcal{R} = \mathcal{P}^{(1)}=\mathcal{P}^{*}$
in \eqref{eq:RG-J2}.
 It is usefull to split $\mathcal{P}^{(2)}(\zeta,\Gamma)$ into a 
 $\Lambda$ independent part $\mathcal{P}^{(2)}_i(z,\Gamma)$ and $\mathcal{P}^{(2)}_\Lambda (z,\Gamma)$. By denoting 
 $p(z, \Gamma)$
 the Laplace transform in $\zeta$ of $\mathcal{P}^{(2)}(\zeta,\Gamma)$, we find that 
 for $z$ and $\Gamma$ finite and fixed, 
$p_\Lambda (z,\Gamma) \to 0$ when $\Lambda \to \infty$.
 Then we  show that the norm of the two parts of the solution satisfy : 
$
 \| \mathcal{P}^{(2)}_i \|_{\zeta} = 
 1-  \| \mathcal{P}^{(2)}_\Lambda \|_{\zeta} = 
 \lim_{z \rightarrow 0} p_i(z,\Gamma) = \Gamma_0 / \Gamma, 
 $
 corresponding to a constant ``decrease'' of the couplings $J^{(2)}$.  
In this regime, the system ``forgets'' its initial conditions and
flows to a general state gouverned by $\mathcal{P}^{(2)}_\Lambda$.
Consistency of conditions {\it (i)} to {\it (iii)} can also easily being checked 
from the properties of the Laplace transform. 
As a consequence, such small next nearest neighbor couplings 
 correspond to an ``irrelevant perturbation'' at this infinite disorder 
fixed point.

 
\begin{figure}[th]
\includegraphics[width=8cm]{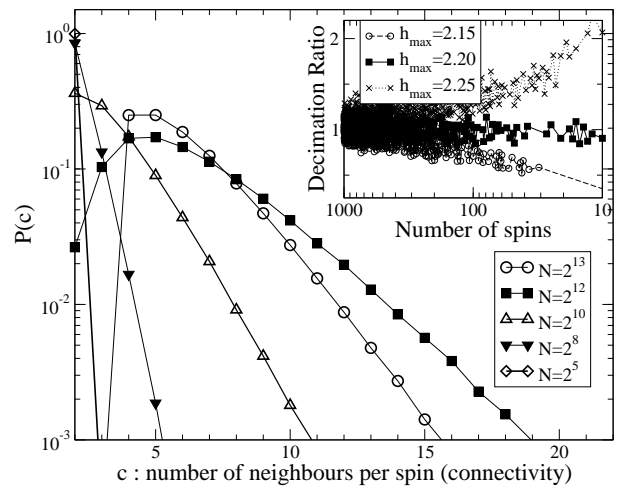}
\caption{\label{fig:connectivity}
Scaling behavior of the distribution of connectivity $P(c)$ for different number $N$ 
of remaining clusters  (spins), for the $J_{1}-J_{2}$ model. The initial size is $N_{0}=2^{14}=16384$ spins, 
and the decimation was performed over $1000$ samples. After a transient regime characterized 
by an algebraic distribution of connectivity, the distribution ultimately renormalizes towards  a delta function 
$c=2$ corresponding to the topology of the chain. 
The inset show the ratio of number of decimated fields over the number of decimated bonds as a function 
of the number of remaining clusters $N$ (see text). The critical point is evaluated as $h_{max}=2.20 \pm 0.05$. }
\end{figure}
 
 To go beyond this perturbative analysis, we have studied the scaling behavior of the zig-zag ladder by 
 implementing numerically the above renormalization rules. We start by choosing a random configuration 
 of fields $h_{i}$ and couplings $J^{(1)}_{i,i+1},J^{(2)}_{i,i+2}$ according to the previous initial distributions probabilities. 
 Then at each step, the energy scale is lowered and the number $N$ of spins is reduced by $1$ according 
 to the decimations rules specified above. This process is continued up to the last remaining spin, and 
 repeated for a number $R=10^{3}$ configurations.  
 No assumption is made on the topology of the renormalized lattice, 
 and we keep {\it a priori} all generated couplings. However, for practical reasons it appears necessary to 
 restrict ourselves to energies larger than a lower cut-off $\Omega_{min}$.  
 With this procedure, the distributions 
 $\mathcal{P}(\zeta,\Gamma),\mathcal{R}(\beta,\Gamma)$ are correctly sampled below 
 $\Gamma_{max}=\ln (\Omega_{0}/\Omega_{min})$ \cite{Motrunich2000}.   For most of our results, this 
 cut-off $\Omega_{min}$ was maintained to negligible values, without any noticeable incidence on the results. 
  For fixed $J^{(2)}_{max}$, the transition is reached by varying the maximum amplitude $h_{max}$ of the 
  fields. We locate a putative infinite disorder phase transition by using the analogy 
  with percolation \cite{Monthus1997}.  At each decimation step $i$, corresponding to a system size $N_{0}-i$, 
  we consider the number of realizations $n_{h}(i)$ where a field was decimated at step $i$, the similarly for 
  the bonds $n_{J}(i)$. At the transition, the ratio $n_{h}(i)/n_{J}(i)$ should become scale invariant, whereas it should 
  diverge or decrease to zero respectively in the disordered or ordered random phases. Moreover, the scaling 
  behavior of this ratio is an excellent way to check for possible finite size effects respective to the topology of the 
  initial lattice. The inset of the figure \ref{fig:connectivity} shows this scaling of the decimation ratio for two values 
  around the candidate critical value of $h_{max}$. Once such candidates for the transition are determined, we 
  have studied the scaling behavior of the distributions functions 
  $\mathcal{P}(\zeta,\Gamma),\mathcal{R}(\beta,\Gamma)$, of the distribution of magnetization $\mu(\Gamma)$, 
  and number of active spins $n(\Gamma)$ in the clusters. This allows to characterize the criticality of the 
  infinite disorder fixed point. Moreover, to fully characterize an infinite disorder fixed point beyond the simple chain, 
  one should also be able to determine the renormalized topology of the critical lattice, and the associated correlations 
  with the couplings.  In a first attempt to study the scaling of this topology, we have followed the distribution of 
  the connectivity of the lattice as the decimation goes on. The results, depicted on fig. \ref{fig:connectivity}, shows 
  that while initially all sites have only $4$ neighbors, the distribution $P(c)$ flows towards an intermediate algebraic 
  distribution at intermediates sizes. While highly connected sites appear, we find by varying our lower cut-off 
  $\Gamma_{max}$ that rather strong correlations exist between the bonds connecting these sites. And while 
  the decimation is pursued, the distribution narrow back towards a delta function peaked on $c=2$, {\it i.e} 
  the lattice is ultimately renormalized towards a chain. We thus find that for the $J_{1}-J_{2}$ model, the infinite 
  disorder fixed point is always given by the fixed point of the chain (see above and \cite{Fisher1995}),
   in agreement with previous results on the similar ferromagnetic two-leg ladder \cite{Rieger2000}
\begin{figure}[th]
\includegraphics[width=8cm]{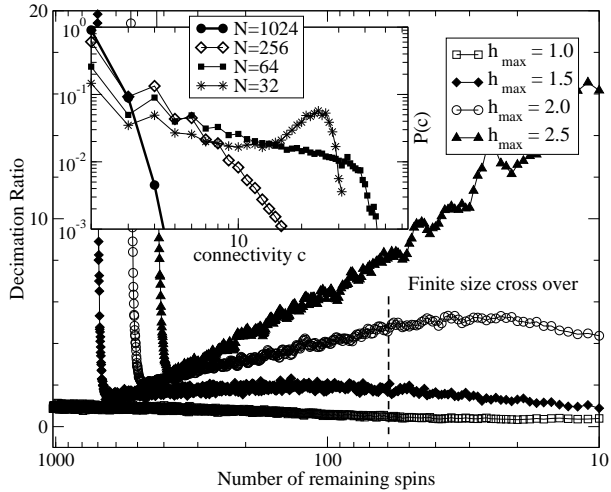}
\caption{\label{fig:SmallWorld}
Decimation ratio for the disordered quantum Ising model on a small world 
lattice. The initial size is $N_{0}=1024$, $q=0.1$ and 5000 samples were used. 
The scaling behaviour of the distribution of connectivity for $h_{max}=1.5$, in the inset, 
shows a broadening of this distribution up to some clear finite size topological effects 
inherent to small world models. 
}
\end{figure}

 The previous results motivated the study of the opposite limit of long-range couplings competing 
 with the initial couplings of the chain. Thus we naturally consider the hamiltonian (\ref{eq:randomIsing}) 
 on a small world lattice\cite{Albert2002}, where beyond the previous nearest neighbors couplings 
 $J^{(1)}_{i,i+1}$, we add random infinite range couplings $J_{i, j}^{LR}$ between any two non-neighbor 
 sites $i$ and $j$, with density $p/N$.  In this paper, the existing 
 couplings $J_{i, j}^{LR}$ and $J^{(1)}_{i,i+1}$ are distributed with the same uniform distribution between 
 $0$ ad $1$.  With these conventions, the average initial connectivity of this lattice is $2+q$. 
  Results of the same numerical decimation procedure as above indicate a phase transition different from 
the previous one (Zig-Zag ladder). In particular, contrarily to the previous case, the distribution of connectivity 
of the renormalized lattice broadens without limit up to some finite size effects. This cross-over happens 
 when the numerical upper bound of the renormalized distribution $P(c)$ becomes of the order of the system size. 
  Once this happens, highly connected sites proliferate, leading to a mean-field like behavior. 


In this paper we have shown how the presence of random signs and further neighbor couplings
affect the critical behavior of the random quantum Ising chain. We have particularly focused on the 
topological properties of the renormalized lattice, and we have explicitly shown 
 how the presence of second neighbors couplings (Zig-Zag ladder) leads to an asymptotic lattice equivalent 
 to a simple chain, proving the irrelevance of the second neighbor couplings perturbation at the infinite disorder 
 fixed point of the chain. On the other hand, the results of our numerical renormalization approach show that 
 the inclusion of an arbitrary density of long range couplings in the chain modifies the scaling behavior 
 of the lattice's topology, and thus the associated critical behavior. 
  These results stress the importance of determining the renormalized topological properties  at any possible 
  infinite disorder transition  beyond the one-dimensional examples. In particular, the intermediate regime 
  we have identified in our study of the Zig-Zag ladder opens the possibility of new infinite disorder scenarii
   for models with correlated long-range couplings.  A natural extension of the present work would 
   certainly focus on random algebraic interactions and the effect of the dimension, possibly relevant to 
   the understanding of the dipolar glass LiHo$_{x}$Y$_{1-x}$F$_{4}$ in a transverse field \cite{Aeppli1987}.

D. Carpentier and P. Pujol would like to acknowledge F. Leonforte for collaboration in a preliminary
investigation related to the present work.

\end{document}